\documentclass[prl,aps,twocolumn,preprintnumbers,floatfix]{revtex4}

\usepackage[mathscr]{eucal}
\usepackage[latin1]{inputenc}
\usepackage{amsmath}
\usepackage{amsfonts}
\usepackage{amssymb}
\usepackage{pictex}
\usepackage{graphicx}

\def\be{\nopagebreak[3]\begin{equation}}
\def\ee{\end{equation}}
\def\ba{\nopagebreak[3]\begin{eqnarray}}
\def\ea{\end{eqnarray}}

\newcommand{\teta}{\rlap{\lower2ex\hbox{$\,\tilde{}$}}\eta{}}

\def\rcr{\rho_c}

\newcommand{\f}{\frac}

\def\f{\frac}

\usepackage{enumerate}

\usepackage{colordvi}
\newcounter{mnotecount}[section]

\newcommand{\comment}[1]{}

\def\f{\frac}

\def\br{{\cal R}}

\begin{document}
\preprint{\vbox{\baselineskip=12pt \rightline{PI-QG-83}}}
\title{Covariant Effective Action for Loop Quantum Cosmology \`{a} la Palatini}

\author{Gonzalo J. Olmo$^{a,b}$ and
Parampreet Singh$^{a}$}
\affiliation{ $^a$Perimeter Institute for Theoretical Physics,
31 Caroline Street North, Waterloo, Ontario N2L 2Y5, Canada\\ 
$^b$Instituto de Estructura de la Materia, CSIC, Serrano 121, 28006 Madrid, Spain}

\begin{abstract}
In loop quantum cosmology, non-perturbative quantum gravity effects lead to
the resolution of the big bang singularity by a quantum bounce {\it without} introducing 
any new degrees of freedom. Though fundamentally discrete, the theory admits a continuum
description in terms of an effective Hamiltonian. Here we provide an algorithm to obtain 
the corresponding effective action, establishing in this way the covariance of the theory for the first time. 
This result provides new insights on the continuum properties of the discrete structure of quantum geometry
and opens new avenues to extract physical predictions such as those related to gauge invariant cosmological perturbations.
\end{abstract}

\maketitle

Understanding the nature of gravity and spacetime at high energies is one of 
the most interesting open issues in theoretical physics. In it 
lie the answers to various questions which Einstein's 
theory of general relativity (GR) fails to address, such as the origin of our 
Universe and the resolution of the big bang singularity. This is also deeply connected with  our  
understanding of the way various dynamical and structural properties of 
the spacetime and the field equations, such as covariance, emerge from a more fundamental 
description. \\
\indent It is generally believed that limitations of GR would be overcome in a 
quantum theory of gravity, which is expected to cure the big bang 
singularity and provide modifications to the Friedman dynamics in the early universe. 
An approach in this direction 
is to find a   renormalizable perturbative theory of quantum 
gravity which agrees with GR at low energies.
This inspired modifications of the 
 Einstein-Hilbert action via addition of terms involving 
higher curvature invariants and higher derivatives of the metric, motivating ansatzes to potentially tame the  initial singularity  (see for example \cite{bran-mukh}).
They inevitably have more degrees of freedom than GR and often 
face limitations such as lack of unitarity, ghosts, and instabilities. 
These effective theories are based on a classical continuum spacetime and are 
 covariant by construction. \\
\indent To faithfully capture the dynamical nature of spacetime, however, we need to 
go beyond the perturbative methods. One such approach is loop quantum gravity, 
which is background independent and non-perturbative \cite{lqg}.
 It is a canonical quantization of gravity  with classical phase space 
 given by the Ashtekar variables: the connection $A^i_a$ and 
the triad $E^a_i$. A key prediction of the theory is the discreteness of 
the eigenvalues of geometrical operators such as volume and area. Thus, the classical 
notion of a smooth differentiable geometry is replaced by a discrete 
quantum geometry. Techniques of LQG have been successfully applied to 
formulate loop quantum cosmology (LQC) which is a non-perturbative quantization 
of cosmological spacetimes \cite{lqc}. In  recent years, extensive analytical 
work and numerical simulations have shown that the big bang singularity 
can be resolved in LQC. The non-perturbative quantum geometric effects 
result in a quantum bounce to a pre-big bang branch when the energy density of the universe reaches 
close to the Planck scale \cite{aps}. Further, analysis from exactly solvable models 
show that the bounce is generic \cite{slqc}.\\
\indent Though the fundamental description in LQC is discrete, it is interesting 
to note that it admits an effective continuum spacetime description which 
successfully captures the quantum gravity effects at high energies and becomes classical at low energies. It is derived from an 
effective Hamiltonian obtained using coherent state techniques. The resulting 
equations of motion yield trajectories which are in excellent 
agreement with the quantum expectation values for the states corresponding 
to realistic universes \cite{vt}. As expected, these non-singular trajectories 
do not exactly follow classical GR but correspond to a modified Friedman dynamics
leading to a bounce at the value of the energy density predicted by the quantum 
theory and recovering classical GR at late times. \\

\indent 
These features and success of LQC allow us to pose questions about aspects which were 
previously poorly understood or unknown. One of such questions is:
How do the classical properties of spacetime change when quantum gravitational 
effects become important? A related question  often posed for any 
canonical quantization is the fate of spacetime covariance. 
If the fundamental picture is discrete, this issue becomes trickier. 
Since in LQC an effective continuum description is available, this question 
can be posed and it is pertinent to ask: Does the effective dynamics of 
LQC which results in a non-singular evolution correspond to a covariant description?
Note that even though, LQC is a quantization only of cosmological spacetimes, 
it is one of the few settings in 3+1 dimensions where a non-perturbative 
quantization has been completely performed and physics beyond  classical 
gravity is well understood. Thus, this query holds promise in providing us with a 
better understanding of at least some of the quantum features of spacetime.

One of the ways to verify if the theory is covariant is to show that it can 
be derived from a (covariant) action. Before we find this action for LQC, let 
us first note one of its interesting features. That is, the  effective 
dynamics of LQC  resolves the big bang singularity {\it without} the
introduction of any new degrees of freedom. It means that the corresponding 
modified  Friedman dynamical equations are second order 
in time, as in GR. This is in contrast with the conventional action based 
perturbative treatments. Here it is important to recall that 
requiring second-order equations and covariance one is uniquely led to the 
Einstein-Hilbert lagrangian density (modulo a cosmological constant) and 
hence to the Einstein field equations. With these apparent tensions, proving the 
covariance of the effective dynamics of LQC comes as a challenging task requiring 
key new insights.

A way out of these problems starts by noting that in formulations with 
actions involving higher order derivatives of the metric one assumes the 
compatibility of the spacetime connection with the metric. In LQG, the 
Ashtekar-Barbero connection is not a spacetime connection \cite{samuel}. 
Further, there exists no corresponding connection operator in the quantum theory. 
It is thus conceivable that the process of loop quantization takes us beyond the compatibility 
condition between the connection and the metric, changing the description of 
spacetime in a fundamental way. Hence, when looking for an effective covariant 
action for LQC, there is no reason to assume any a priori relationship between 
the spacetime connection and the metric.

If in the gravitational action metric and connection are treated as independent fields one deals with a
metric-affine theory. If the connection is torsionless and uncoupled to matter then one arrives at the 
Palatini formulation of gravity. In this latter approach particles follow geodesics of the Levi-Civita 
connection of the metric rather than those of the independent connection.
Only for the Einstein-Hilbert lagrangian do metric and Palatini formalisms
lead to the same dynamics. In general, they are completely different theories. 
Unlike the metric formulation, where higher order terms motivated from perturbative 
techniques in quantum gravity have been well studied, so far there is little  
``inspiration'' from a fundamental theory to consider higher curvature actions in the 
Palatini framework. Phenomenological investigations of Palatini theories have recently 
gained some attention in relation with the late time cosmic acceleration \cite{mod-grav-long}. Though 
catastrophic matter instabilities have been found regarding Palatini models with infrared 
modifications of gravity \cite{Sot08,Olmo07,Fla04}, it is also true that models with ultraviolet modifications 
are as robust as GR  within the experimental limits for suitable choices of model parameters \cite{Olmo08a,Olmo08b}. 

The most general Palatini action can be represented as a function of the form 
$f(\br, \br_{\mu \nu} \br^{\mu\nu}, \br_{\alpha \beta \gamma \delta}  \br^{\alpha \beta \gamma \delta}, \ldots)$, where one may include covariant derivatives of functions of the metric and derivatives of curvature invariants. Here 
$\br$ denotes the Riemann curvature of the independent connection. In general, the field equations of these theories have the same number of degrees of freedom as GR. This is due to the fact that the independent connection satisfies a constraint equation, rather than a dynamical evolution equation, whose solution can be expressed as the Levi-Civita connection of an auxiliary metric related to the spacetime metric and the energy-momentum tensor of matter. In the simplest case in which the action is of the form $f(\br)$, the auxiliary and spacetime metrics are conformally related, with the conformal factor being a function of the trace of the energy-momentum tensor of matter. 
The connection can then be readily solved in terms of the spacetime metric and the matter and eliminated from the field equations. 
Though the resulting theory has the same configuration space as GR, its dynamics is different. The role of the Palatini lagrangian $f(\br)$ is just to change the way matter generates the spacetime curvature, i.e., 
it modifies the GR relation $R=-\kappa^2T$ to arbitrary $f(\br)$ lagrangians [see (\ref{eq:Pal-trace}) below]. This is to be contrasted with the metric formulation, where the lagrangian $f(R)$ turns the scalar curvature $R$ into a dynamical entity 
which satisfies a second-order differential equation \cite{Olmo05} and, therefore, the theory has higher degrees of freedom.

The kinematical similitudes between LQC and Palatini theories raise a compelling question: 
Can the effective dynamics of  LQC be expressed in the form of an effective
Palatini theory? If the answer is positive then apart from establishing covariance of the modified Friedman dynamics in LQC and gaining insights on 
effects of quantum gravity on the properties of continuum spacetime, a  multitude of 
benefits result. We will be able to understand the way non-perturbative canonical quantum gravity effects can be 
captured in the effective action treatments. Further, many interesting questions which are beyond the 
scope or are difficult to address in conventional Hamiltonian treatments could be posed and answered. \\

Let us consider a flat isotropic and homogeneous FRW spacetime sourced with a 
massless scalar field $\phi$ with canonical momentum $p_\phi$ (satisfying $\{\phi,p_\phi\} = 1$). 
This model has been successfully quantized in LQC 
and its physics has been well understood  \cite{aps}. The underlying quantum 
constraint is non-local and uniformly discrete in volume. An effective description of the quantum dynamics can be obtained using geometric methods where  one treats the Hilbert space as an infinite dimensional 
 quantum phase space with a fiber bundle structure. Using coherent states, an approximately  horizontal 
section which is preserved under the quantum Hamiltonian flow to a desired accuracy can be obtained \cite{vt}.
The resulting effective Hamiltonian (or the modified Friedman dynamics) describes the underlying quantum 
evolution extremely well at all scales for universes which grow to a macroscopic size  and leads to a 
rich phenomenology (see for eg. \cite{svv}).

The modified Friedman equation resulting from the effective Hamiltonian methods in LQC (for a massless scalar) is \cite{vt}:
\be
3H^2=\kappa^2\rho\left(1-\frac{\rho}{\rcr}\right), ~ \rcr := \f{\sqrt{3}}{16 
\pi^2 \gamma^3 G^2 \hbar}\label{eq:H2-LQC} ~
\ee
where $\kappa^2 = 8 \pi G$.
Since, the loop quantization does not affect the matter Hamiltonian, the  
Klein-Gordon equation and the conservation law are unmodified, i.e. $\dot{\rho}=-3H(\rho+P)$ (where $\rho$ and $P$ are energy density and  pressure of the scalar field respectively).
It is straightforward to see that when $\rho \ll \rcr$, 
the modified Friedman dynamics reduces to the classical equations. 
At $\rho = \rcr$,
the Hubble rate vanishes and also $\ddot a > 0$, implying the bounce of the universe. The occurrence of the bounce is purely a quantum gravitational effect which 
disappears when $G \hbar$ vanishes (implying divergence of $\rcr$).

To obtain the modified Friedman dynamics of LQC from a covariant 
Palatini action framework, our approach will be to solve the inverse problem --  find the Lagrangian, given the 
equations of motion. Such an action will be effective in the sense that it provides a covariant 
description of LQC dynamics as obtained from the effective Hamiltonian \cite{vt}. 
A generalized Palatini action is given by 
\be
S(g,\Gamma,\phi) = \f{1}{2 \kappa^2} \hskip-0.1cm\int \hskip-0.1cm d^4 x \sqrt{-g} f({\cal R}, \br_{\mu \nu} \br^{\mu\nu}, ...) + S_{\mathrm{matt}}(g_{\mu \nu},\phi)
\ee
with ${\cal R} := g^{\mu \nu} {\cal R}_{\mu \nu}(\Gamma)$. For simplicity we consider the gravitational part 
only as a function  $f(\br)$. Its variation with respect to the metric and connection results in 
\begin{eqnarray}\label{eq:met-var-P}
f'(\br)\br_{\mu\nu}(\Gamma)-\frac{1}{2}f(\br)g_{\mu\nu}&=&\kappa ^2T_{\mu\nu } \\
\nabla^\Gamma_{\mu}\left[\sqrt{-g}f'(\br)g^{\alpha\beta}\right]&=&0 \label{eq:con-var-P} 
\end{eqnarray}
where the prime denotes derivative with respect to ${\cal R}$. The covariant derivative $\nabla^\Gamma_{\mu}$ is not compatible
with the metric: $\nabla^\Gamma_\mu g_{\alpha\beta}\neq0$. However, it satisfies $\nabla^\Gamma_\mu t_{\alpha\beta}=0$ where $t_{\mu\nu}=f' g_{\mu\nu}$. 
Thus, $\Gamma$ is  the Levi-Civita connection of the auxiliary
metric $t_{\mu \nu}$. 
The trace of Eq. (\ref{eq:met-var-P}) leads to a generalization of the algebraic relation $R = - \kappa^2 T$ for non-linear $f(\br)$ in Palatini:
\begin{equation}\label{eq:Pal-trace}
\br f'(\br)-2f(\br)=\kappa^2 T ~.
\end{equation}
This algebraic equation can be solved to obtain ${\cal R} = \br(T)$. 
Inserting the solution for the connection, in terms of $f'(\br(T))$ and $g_{\mu\nu}$, in (\ref{eq:met-var-P}) one finds
\begin{eqnarray}\label{neweinstein}
G_{\mu \nu}(g) &=& \nonumber \f{\kappa^2}{f'} T_{\mu \nu} - \f{\br f' - f}{2 f'} g_{\mu \nu} + \f{1}{f'}\left(\nabla_\mu \nabla_\nu f' - g_{\mu \nu} \Box f'\right)\\ 
&& - \f{3}{2 f'^2} \left(\partial_\mu f' \partial_\nu f' - \f{1}{2} g_{\mu \nu} (\partial f')^2\right) 
\end{eqnarray}
where $G_{\mu \nu}(g) := R_{\mu \nu}(g) - \tfrac{1}{2} g_{\mu \nu} R(g)$. 
Note that Eqs.(\ref{eq:met-var-P}) and (\ref{neweinstein}) are conformally related and the latter implies the former. Further, the conservation law is unmodified, i.e. $\nabla_{\mu} T^{\mu \nu} = 0$. Note also that in vacuum we find $G_{\mu \nu}=-\Lambda g_{\mu\nu}$, with $\Lambda=\f{\br f' - f}{2 f'}$ evaluated at $T=0$, which recovers the dynamics of GR plus a cosmological constant. This guarantees that the Cauchy problem in vacuum is well-posed (the opposite, however, has been claimed in \cite{LTF07}). From (\ref{neweinstein}), 
the modified Friedman equation becomes
\begin{equation}\label{eq:H2-Pal}
3H^2=\frac{f'[2\kappa^2 \rho+\br f'-f]}{2\left(f'+\frac{f''}{2}\frac{\dot{\br}}{H}\right)^2}
\end{equation}
where $\dot {\br}/H = - 12 \, \kappa^2 \rho/(\br f'' - f')$.
It is to be emphasized that since $f(\br)$ is a function of $T$, the right hand side of (\ref{eq:H2-Pal}) does not involve any higher
derivatives of geometrical quantities and is just a function of the matter sources.
The problem of finding the effective action for the LQC dynamics thus 
reduces to finding an $f(\br)$ satisfying
\begin{equation}\label{eq:Pal=LQC}
\frac{f'[2\kappa^2 \rho+\br f'-f]}{2\left(f'+\frac{f''}{2}\frac{\dot{\br}}{H}\right)^2}=\kappa^2\rho\left(1-\frac{\rho}{\rho_c}\right)
\end{equation}
where $\rho=\rho(\br)$ is a solution to (\ref{eq:Pal-trace}). This is just a second-order differential equation for $f(\br)$
\begin{equation}\label{eq:ode-f}
f''=-f'\left(\frac{f' A-B}{2\left(\br f' - 3 f\right) A+ \br B}\right)
\end{equation}
where $A = [2(\br f'-2f)(2\br_c-(\br f'-2f))]^{1/2}$ and $B =  2 [\br_c f'(2\br f'-3 f)]^{1/2}$ 
and  $\br_c\equiv \kappa^2\rho_c$. Physically acceptable solutions should be free of singularities and have the property that the function $\rho(\br)$ maps the full range of values $\rho \in [0,\rho_c]$. These conditions are equivalent to demanding that the acceleration $\ddot{a}$ at the bounce, $\dot{a}=0$, be the same for both LQC and Palatini $f(R)$, which implies that the bounce must occur at $R=-12 \br_c$, where $f'\to 0$ and $\rho\to\rho_c$. 
Numerically we find a family of solutions which converge to a unique function satisfying the above constraints. This shows that {\it  a physically consistent $f(\br)$ solution to (\ref{eq:ode-f}) corresponding to the effective dynamics of LQC exists.} Furthermore, the inverse problem has a unique solution. It is important to note that at curvatures $|\br| \ll \br_c$,  $f'(\br)$ approximates unity, i.e., 
$f(\br) \approx \br$, and the solution leads to the classical Friedman dynamics.

An analytical form for $f(\br)$ can be obtained by means of interpolation techniques. At low curvatures the numerical solution can be approximated via 
\be\label{fr-l}
f(\br) = -\int d\br \tanh\left(\frac{5}{103}\ln\left[\left(\frac{\br}{12\br_c}\right)^2\right]\right) ~.
\ee
An interesting function which captures the
loop quantum dynamics from sufficiently low to the maximum value of the curvature is
\be\label{frh}
f(\br)=\frac{\br}{12}\left(1-\frac{1}{2}\ln\left[\left(\frac{\br}{12\br_c}\right)^2\right]\right)+\frac{\br(\br+12\br_c)^2}{6500 \br_c^2} ~.
\ee
The first term dominates at higher curvatures and incorporates the non-perturbative quantum gravity effects that lead to the cosmic bounce in LQC when $\rho = \rcr$ (see Fig. 1). 
Note that in contrast with the conventional metric formulation, which generally incorporates perturbative quantum gravity effects via a finite number of terms, the above analytical forms are infinite series. This distinction primarily arises because in non-perturbative LQC quantum geometry effects are non-local. If the fundamental theory were local, a finite number of terms would have sufficed.

Our investigation to find a covariant action for the effective dynamics of 
LQC provides a much needed motivation from a fundamental description to 
study $f(\br)$ modifications of gravity and its possible extensions 
in the Palatini formalism. The covariant action we find here 
leads to non-singular isotropic cosmological dynamics mimicking that of LQC. 
Based on this action, generalizations to other cosmologies and black hole 
spacetimes can be considered, which opens a rich avenue to study non-singular 
spacetimes in the Palatini approach. Further, the availability of an action 
framework opens a straightforward way to perform a gauge invariant
investigation of cosmological perturbations in loop cosmology which is 
very important to extract physical predictions from the theory.

Let us now revisit the  primary question we posed in this letter, namely, 
the way properties of classical spacetime are affected in a quantum theory of 
gravity. We considered here as an example non-perturbative loop quantization 
of cosmological spacetimes. The underlying geometry at the fundamental level is 
discrete, however the theory admits an effective continuum spacetime. 
The resulting dynamics, though non-singular, was never established to be 
covariant until now. Demanding its covariance, we find that  
the connection must be regarded as independent of the metric in the derivation of the field equations. Though this procedure might not be the only solution to the problem considered here, it provides 
new insights on the kind of fields that an action should contain to capture non-perturbative quantum
gravity effects. This suggests that, unlike in the classical spacetime of GR, the metric might not be 
the sole fundamental geometric entity. The role of the independent connection is, however, unconventional 
in the sense that it satisfies a constraint equation rather than a dynamical 
second-order, differential equation. This, in turn, is what guarantees the kinematical
similitudes between LQC and Palatini theories. The solution to the constraint shows
that matter and geometry get entangled in a non-trivial way with important consequences. 
In fact, it turns out that the spacetime metric depends on the 
local energy-momentum densities, which leads to strong (and unacceptable) backreaction effects in infrared-corrected 
models \cite{Olmo08a} but removes the big bang singularity in models with appropriate
ultraviolet corrections, as shown here. 

Though surprising at first, departures from a purely metric-based framework have been often 
considered as a necessary requirement if we wish to overcome the 
limitations of GR, such as non-renormalizability \cite{krasnov}. It has also been 
argued that if the fundamental description is discrete, like in the crystalline 
structure of solids, then the metric alone is insufficient  to capture all the 
geometric properties and the effective continuum spacetime may be non-Riemannian \cite{hehl}. 
As in a Bravais lattice, the underlying structure in LQC is discrete and our 
results show that its effective continuum spacetime indeed takes us beyond metric 
properties. This holds similarity with investigations on studies of continuum properties of 
crystals \cite{kroner}. 

To summarize, we have shown for the first time that despite apparent
tensions with the conventional wisdom based on perturbative ideas (in
metric formalism), a covariant effective action which reproduces the
dynamics of LQC exists in a framework in which metric and connection are
regarded as independent. 

\begin{figure}[tbh!]
\begin{center}
\includegraphics[scale=0.45]{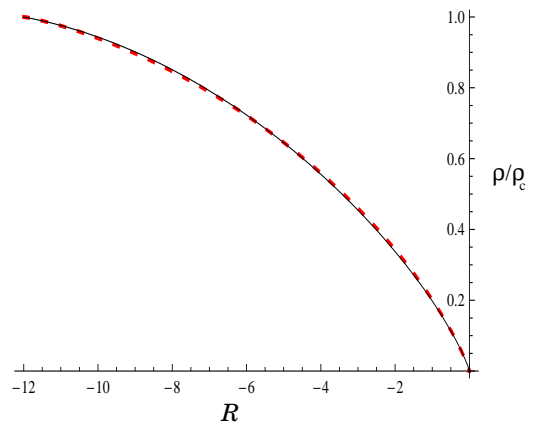}
\caption{The energy density in units of $\rcr$ is plotted for the numerical solution (dashed line) and the $f(\br)$ in (\ref{frh}) (solid line). At $\br = - 12 \br_c$, both of them yield $\rho = \rcr$ leading to a non-singular bounce.\label{Fig:den}}
\end{center}
\end{figure}

Loop quantization, at least for the simplest cosmological models,
seems to take us away from metric-connection compatibility, thus
allowing reconciliation between the lack of new degrees of freedom in
the modified Friedman dynamics and covariance.  
It remains to be seen whether these novel features survive a more general
quantization. Further improvements in the approximations used in obtaining the 
effective Hamiltonian of LQC \cite{vt} and its generalization to include corrections such as those 
originating from quantum properties of state and to the anisotropic and 
inhomogeneous spacetime would result in further insights on the continuum 
properties of the discrete structure of quantum geometry.
Our analysis should be seen as a first step towards 
such explorations, which provide a glimpse of new ways in which matter and 
geometry might get entangled, via an independent connection, and the nature of the effective
spacetime emergent from a quantum geometry.

\begin{acknowledgments}
We are grateful to A. Ashtekar, A. Corichi and B. Dittrich for useful discussions and extensive comments. We thank F. Barbero, R. Maartens, G. Mena-Marugán and K. Vandersloot for comments.
PS also thanks  T. Biswas and K. Vandersloot for discussions on metric formulation of the problem.
GO has been supported by MICINN. 
Research of PS is supported by Perimeter Institute for Theoretical Physics.  
Research at Perimeter Institute is supported by the Government
of Canada through Industry Canada and by the Province of Ontario through
the Ministry of Research \& Innovation.
\end{acknowledgments}

\end{document}